\documentclass[aip,apl,twocolumn,reprint,amsmath,amssymb,superscriptaddress]{revtex4-1}
\usepackage[utf8]{inputenc}
\usepackage{graphicx}

\begin{document}
% Title Page
\title{Single Hole Transport in a Silicon Metal-Oxide-Semiconductor Quantum Dot}
\author{R. Li}
\affiliation{School of Physics, University of New South Wales, Sydney NSW 2052, Australia}

\author{F. E. Hudson}
\author{A. S. Dzurak}
\affiliation{Australian National Fabrication Facility, University of New South Wales, Sydney NSW 2052, Australia}
\affiliation{Centre of Excellence for Quantum Computation and Communication Technology, School of Electrical Engineering and Telecommunications, University of New South Wales, Sydney NSW 2052, Australia}

\author{A. R. Hamilton}
\email{Alex.Hamilton@unsw.edu.au}
\affiliation{School of Physics, University of New South Wales, Sydney NSW 2052, Australia}

\date{\today}

\begin{abstract}
We describe a planar silicon metal-oxide-semiconductor (MOS) based single hole transistor, which is compatible with conventional Si CMOS fabrication. A multi-layer gate design gives independent control of the carrier density in the dot and reservoirs. Clear Coulomb blockade oscillations are observed, and source-drain biasing measurements show that it is possible to deplete the dot down to the few hole regime, with excited states clearly visible. The architecture is sufficiently versatile that a second hole dot could be induced adjacent to the first one.
\end{abstract}

\maketitle

%--------------background and paper overview---------------------
%\section{Introduction}
Over the past 15 years much effort has gone into the development and study of electron quantum dots as artificial atoms~\cite{0034-4885-64-6-201, 0957-4484-22-33-335704}, ultra-sensitive electrometers~\cite{PhysRevB.81.161308}, and quantum bits~\cite{Petta30092005} for quantum information applications. To use an electron in a quantum dot as a spin qubit requires long spin life-time T$_1$ and coherence-time T$_2$~\cite{PhysRevA.57.120, RevModPhys.76.323}. Significant progress has been made with III-V semiconductor based devices, although T$_2$ is limited by the hyperfine interaction between the electron spin and nuclei in the host crystal~\cite{RevModPhys.79.1217}. Spin qubits based on group-IV semiconductors have recently shown long T$_1$ and T$_2$ times~\cite{Morello2010,Pla2012}. However even in silicon based devices challenges remain due to the presence of nonzero nuclear spin in isotopes of Si, the valley degree of freedom in conduction band~\cite{PhysRevLett.88.027903}, and disorder at the Si/SiO$_2$ interface.

Recently holes in quantum dots have attracted significant interest~\cite{PhysRevLett.95.076805, Gerardot2008} since the strong spin-orbit coupling enables all electrical spin manipulation~\cite{PhysRevLett.107.176811,PhysRevLett.109.107201}, while the hyperfine interaction between holes and nuclear spins is strongly suppressed~\cite{doi:10.1021/nl201211d}, promising longer T$_2$. Besides, for holes in silicon there is no valley degeneracy. However, to date there have been only a few studies of holes in gate defined quantum dots\cite{dotsch:341,grbic:232108,PhysRevLett.107.076805,Hu2012}. In this letter, we describe a planar silicon metal-oxide-semiconductor (MOS) based single hole transistor, which is compatible with conventional Si CMOS fabrication.

%--------------device layout and operation principle-------------

\begin{figure}[h]
 \centering
 \includegraphics{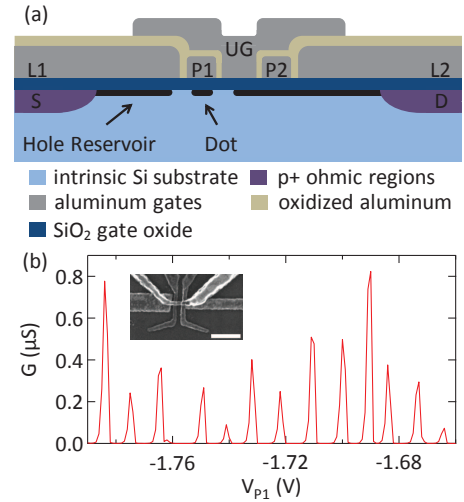}
 \caption{(a) Schematic cross section of the device. The hole reservoirs are induced by lead gates L1 and L2. For the data shown below the dot was induced below plunger gate P1, while UG and P2 were biased to extend the hole reservoir. (b) Conductance vs plunger gate bias, showing periodic Coulomb blockade oscillations in the many hole regime. V$_\mathrm{L1}$=V$_\mathrm{L2}$=V$_\mathrm{P2}$=-4~V, and V$_\mathrm{UG}$=0~V. Inset shows the SEM image of a typical device. The white scale bar is 200~nm.} \label{fig:layout}
\end{figure}

The MOS structure studied in this work was fabricated from a high-resistivity ($\rho$~$>$~10~k$\Omega\cdot$cm) (100) silicon substrate. Field-oxide, boron-diffused ohmic regions, and thin gate-oxide (thickness~$\sim$~5.9~nm) were defined by standard micro-fabrication techniques. Subsequently, multi-level aluminum gates were patterned by electron-beam lithography and lift-off. The gates were insulated from each other by a thin native AlO$\mathrm{x}$ layer~\cite{doi:10.1021/nl070949k}. The final stage was a forming gas anneal to reduce the Si/SiO$_2$ interface trap density and enhance low-temperature performance~\cite{doi:10.1021/nl070949k}.

Figure~\ref{fig:layout}(a) shows a schematic cross section of the device. There are three layers of gates: the first is the two plunger gates (P1 and P2), each 30~nm wide with a separation of 30~nm between them. The middle layer consists of the lead gates (L1 and L2) which were kept at -4~V to induce the source and drain hole reservoirs. The upper gate (UG) has a width of 50~nm and extends over P1 and P2. The multiple gates allow considerable flexibility over device operation. Gates L1 and L2 were always negatively biased to induced holes into the leads, but the remaining gates could either be biased negative to induced holes underneath them, or positive to form tunnel barriers between regions of holes. In the following experiments we used a single gate, P1, to localize holes into a quantum dot and control the dot occupancy. In this mode of operation the entrance and exit tunnel barriers were formed due to the oxidized aluminum layer between different gates, and the upper gate was grounded as it had little effect on the dot. Gate P2 was kept at a large negative bias, to ensure that it was transparent. In this biasing arrangement the lithographic dimensions of the dot were %defined by L1 and P1, and is estimated to be $\sim100\times30$~nm$^2$ \emph{WHY?} .
defined by the width of P1 (30~nm) and the fringing field from L1 (150~nm in width), so we estimate the dot area $\sim3\times10^3$~nm$^2$.

Several devices were tested at 4~K, with a yield of $\sim50$\%. Further measurements were performed on one device in a dilution fridge with base temperature of 30~mK, using standard two-terminal lock-in techniques with a 100~$\mathrm{\mu}$V ac excitation voltage. Fig.~\ref{fig:layout}(b) shows the Coulomb blockade oscillations obtained when sweeping gate P1, demonstrating that the device functions as a single hole transistor. The number of holes in the dot was estimated to be $\sim25$.

%--------------show the dot location-----------------------------

\begin{figure}[h]
 \centering
 \includegraphics{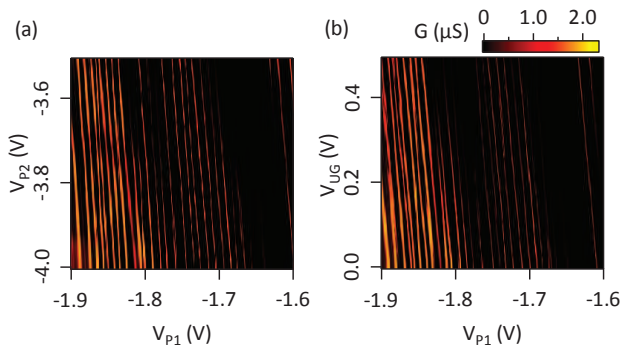}
 \caption{Charge stability diagram of P1 vs. (a) P2 with V$_\mathrm{UG}$=0~V, and (b) UG with V$_\mathrm{P2}$=-4~V. In both plots, V$_\mathrm{L1}$=V$_\mathrm{L2}$=-4~V. The dense parallel lines intercepting V$_\mathrm{P1}$ axis indicates single dot operation and that the dot was strongly coupled to P1.} \label{fig:CSD}
\end{figure}

\begin{table}[ht]
\caption{Gate capacitance to the dot. The capacitance values are estimated from the average line-spacing in charge stability diagrams.}
\centering
\begin{tabular}{l l l l l l}
 \hline\hline
 Gate & P1 & P2 & L1 & L2 & UG\\
 \hline
 C~(aF) & 16.8 & 0.70 & 1.48 & 0.28 & 0.83\\
 \hline
\end{tabular}
\label{table:cap}
\end{table}

Figure~\ref{fig:CSD} shows the conductance of the quantum dot as a function of the biases on gates P1, P2 and UG. These charge stability diagrams show almost vertical lines, indicating that the dot was most strongly coupled to the P1 gate, and the P2 and UG gates had a much weaker capacitive coupling to the dot. Similar results were obtained for the L1 and L2 gates. The large number of periodic oscillations shows that the dot cannot be due to unintentional dopants or defects.  The capacitance of the various gates to the dot was determined from the periodicity of the oscillations in the charge stability diagrams, as shown in Table~\ref{table:cap}. These data confirm that the dot was located under P1, since the capacitance to all other gates was much smaller. We also estimated the size of the dot using a simple parallel plate capacitor model with a silicon oxide thickness $d=5.9$~nm, and obtained an area of $\sim2900$~nm$^2$, in good agreement with the area estimated from the lithographic dimensions of the dot.

%--------------dot in few hole regime----------------------------

%\section{Source-Drain Bias spectroscopy}

\begin{figure}[h]
 \centering
 \includegraphics{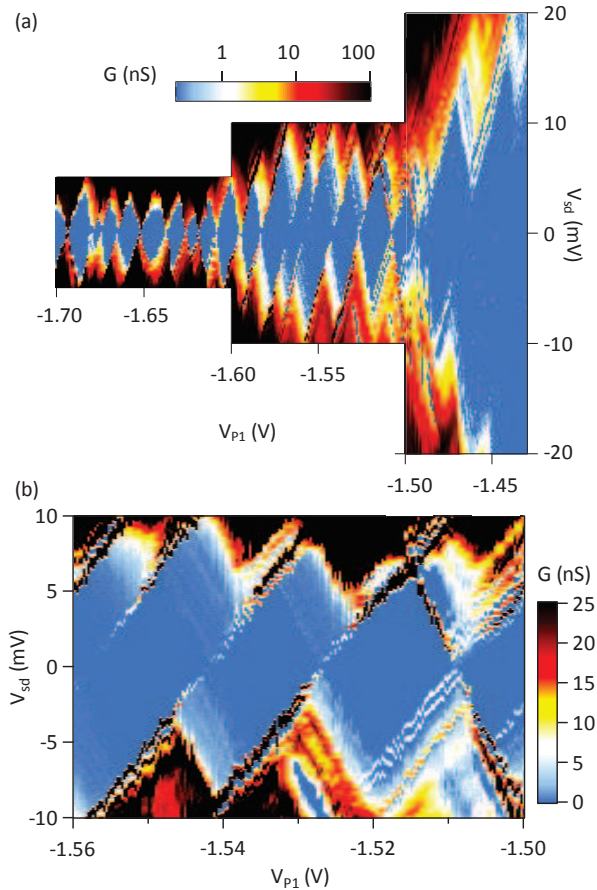}
 \caption{(a) Source-drain biasing of the hole quantum dot, showing Coulomb diamonds down to last few holes. V$_\mathrm{L1}$=V$_\mathrm{L2}$=V$_\mathrm{P2}$=-4~V, and V$_\mathrm{UG}$=0V. (b) A close-up of the Coulomb diamonds, showing that excited states can be resolved.} \label{fig:CBD}
\end{figure}

Figure~\ref{fig:CBD}(a) shows source drain bias spectroscopy measurements, with well resolved Coulomb diamonds. At high gate bias ($|$V$_\mathrm{P1}|$~$>$~1.6~V), the number of holes is greater than ten and the charging energy of the dot was approximately constant at E$_\mathrm{c} \sim5$~meV. As the dot was pinched off, by making V$_\mathrm{P1}$ more positive, the charging energy increased suggesting that the dot was shrinking in size and approaching the few hole limit.  Finally for V$_\mathrm{P1}>-1.5$~V the charging energy increased rapidly and the Coulomb diamonds no longer close.
It is tempting to ascribe this opening of the last Coulomb diamond as signaling the last occupied hole state in the dot. However, the observation of excited states at V$_\mathrm{P1}=-1.47$~V shows that there must be at least one hole in the dot for V$_\mathrm{P1}>-1.47$~V, suggesting the last hole charge state could not be reached in these measurements. No Coulomb diamonds could be resolved for V$_\mathrm{P1}>\sim-1.47$~V, as the device became so pinched off that the conductance dropped below the background noise level of 1~nS (I=0.1pA). However, the $\sim$~10~meV charging energy of the last diamond is a strong indication that we were approaching the last few holes in the dot.

The well defined confining potential of the dot is further highlighted by the slope of the edges of the Coulomb diamonds. The slopes are the same for all diamonds in Fig.~\ref{fig:CBD}(a), giving a lever-arm $\alpha$=C$_\mathrm{P1}/$C$_\mathrm{\Sigma}$=0.36. This suggests that the dot was defined underneath the central region of P1, and was not affected by disorder even in the few hole limit (since $\alpha$ would change and additional features would be observed in the bias spectroscopy if disorder induced parasitic dots were forming). Furthermore the slope of the diamonds allowed the capacitive coupling to the source and drain reservoirs to be estimated, giving C$_\mathrm{S}/$C$_\mathrm{P1}$=1.1 and C$_\mathrm{D}/$C$_\mathrm{P1}$=0.7~\cite{RevModPhys.79.1217}. The dot was more strongly coupled to the  source reservoir than the drain reservoir, consistent with the geometry of the device.

Figure ~\ref{fig:CBD}(b) is a close up of the Coulomb diamonds, showing the excited states of the hole quantum dot. The excited states manifest as thin lines of high-conductance running parallel to the edge of the diamonds outside the Coulomb blockade region.
The spacing of the excited states was $\Delta E \sim800~\mu$eV at V$_\mathrm{P1}=-1.51$~V, although even larger energy spacings $\sim2$~meV could also be resolved. For comparison, measurements of a silicon electron quantum dot fabricated using the same approach and with similar lithographic dimensions showed $\Delta E$ up to 600~$\mu$eV~\cite{doi:10.1021/nl070949k}. Since the hole mass is significantly larger than the electron mass, this would suggest the excited state spacing $\Delta E$ measured for the hole device should be smaller than in Ref.~\onlinecite{doi:10.1021/nl070949k}. However, the hole band structure is more complex than the electron bands, and is further complicated by the lateral confinement in the quantum dot. The thickness of the 2D hole system is $\sim10$~nm, comparable to the length-scale of the in-plane confinement geometry, indicating that the quantization of the hole states should be treated in 3D. The precise nature of the hole states, including the spin properties, shape of the orbital states and the degree of light and heavy hole mixing, is highly sensitive to the confining potential and will be a fruitful area for future research.

%--------------double dot----------------------------------------

\begin{figure}[h]
 \centering
 \includegraphics{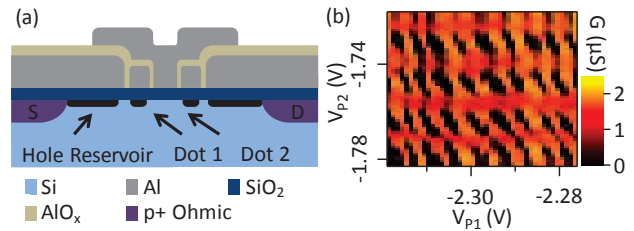}
 \caption{(a) Schematic cross section of the device when the gates are biased to form a double hole quantum dot system. The difference to Fig.~\ref{fig:layout}(a) is that P1 and P2 were in the same bias range and that a dot was induced under each of them. (b) Charge stability diagram obtained by sweeping gates P1 and P2 for V$_\mathrm{L1}$=V$_\mathrm{L2}$=-4~V and V$_\mathrm{UG}$=0~V.}\label{fig:DD}
\end{figure}

Finally we show that this device can also be operated as a hole double quantum dot, by changing the bias on gate P2 so that a second dot formed as sketched in Figure~\ref{fig:DD}(a). The resulting charge stability diagram is presented in Fig.~\ref{fig:DD}(b), where the bias on gate P2 has been reduced from -4~V to a bias  similar to $V_{P1}$.  Dark regions indicate Coulomb blockade where the double dot maintains the same charge configuration, and bright lines indicate current transport through the double dot, demarking regions where the hole occupation changes. In the top right of Fig.~\ref{fig:DD}(b) the vertical lines show that gate P1 controls the hole number in dot 1, while the horizontal lines show that P2 controls the occupancy of the second dot. The horizontal line spacing is approximately twice that of the vertical spacing, suggesting that the capacitance between gate P2 and dot 2 is twice that between gate P1 and dot 1. This is consistent with the lithographic gate dimenions, since the width of gate L1 is twice that of L2 (see SEM image in Fig.~\ref{fig:layout}(b). At more negative $V_{P1}$ and $V_{P2}$ the number of holes in the dots increased and there is evidence of coupling between the dots, as the lines become more diagonal.

%--------------summary-------------------------------------------

In summary, we have fabricated single hole transistors based on a planar silicon MOS structure. A well-defined hole quantum dot could be induced, and operated in both the many-hole and few-hole regimes. Bias spectroscopy measurements show that the device can be operated down to the few hole regime, showing large charging and excited state energies. The flexibility of the multi-gate structure also made it possible to form a second hole dot, with the charge stability diagram displaying weak coupling between the two dots. These devices will allow future studies of individual hole spins in standard silicon MOS structures.

The authors thank N. S. Lai and A. Morello for helpful discussions, A. Rossi for help with fabrication, L. A. Yeoh and A. Srinivasan for help with the dilution refrigerator, and J. Cochrane for technical support.
A.R.H. acknowledge support from the Australian Research Council (DP120102888 and DP120101859).  F.E.H and A.S.D. acknowledge support from the Australian Research Council (CE11E0096) and the U.S. Army Research Office (contract W911NF-13-1-0024). Experimental devices for this study were fabricated with support from the Australian National Fabrication Facility, UNSW.

%--------------note----------------------------------------------
Note added: After completing these measurements we became aware of similar experiments underway elsewhere \cite{2013arXiv:1304.2870}.
%\bibliography{./RefDatabase.bib}

\end{document}